\documentstyle[12pt]{article}

\catcode`\@=11
\long\def\@makefntext#1{
\protect\noindent \hbox to 3.2pt {\hskip-.9pt
$^{{\ninerm\@thefnmark}}$\hfil}#1\hfill}		%CAN BE USED

\def\@makefnmark{\hbox to 0pt{$^{\@thefnmark}$\hss}}  %ORIGINAL
	
\def\ps@myheadings{\let\@mkboth\@gobbletwo
\def\@oddhead{\hbox{}
\rightmark\hfil\ninerm\thepage}
\def\@oddfoot{}\def\@evenhead{\ninerm\thepage\hfil
\leftmark\hbox{}}\def\@evenfoot{}
\def\sectionmark##1{}\def\subsectionmark##1{}}

\newcounter{sectionc}\newcounter{subsectionc}\newcounter{subsubsectionc}
\renewcommand{\section}[1] {\vspace*{0.6cm}\addtocounter{sectionc}{1}
\setcounter{subsectionc}{0}\setcounter{subsubsectionc}{0}\noindent
	{\normalsize\bf\thesectionc. #1}\par\vspace*{0.4cm}}
\renewcommand{\subsection}[1] {\vspace*{0.6cm}\addtocounter{subsectionc}{1}
	\setcounter{subsubsectionc}{0}\noindent
	{\normalsize\it\thesectionc.\thesubsectionc. #1}\par\vspace*{0.4cm}}
\renewcommand{\subsubsection}[1]
{\vspace*{0.6cm}\addtocounter{subsubsectionc}{1}
	\noindent  
{\normalsize\rm\thesectionc.\thesubsectionc.\thesubsubsectionc.
	#1}\par\vspace*{0.4cm}}
\newcommand{\nonumsection}[1] {\vspace*{0.6cm}\noindent{\normalsize\bf #1}
	\par\vspace*{0.4cm}}
					
\newcounter{appendixc}
\newcounter{subappendixc}[appendixc]
\newcounter{subsubappendixc}[subappendixc]

\renewcommand{\appendix}[1] {\vspace*{0.6cm}
        \refstepcounter{appendixc}
        \setcounter{figure}{0}
        \setcounter{table}{0}
        \setcounter{equation}{0}
        \renewcommand{\thefigure}{\Alph{appendixc}.\arabic{figure}}
        \renewcommand{\thetable}{\Alph{appendixc}.\arabic{table}}
        \renewcommand{\theappendixc}{\Alph{appendixc}}
        \renewcommand{\theequation}{\Alph{appendixc}.\arabic{equation}}
        \noindent{\bf Appendix \theappendixc #1}\par\vspace*{0.4cm}}

\def\abstracts#1{{
	\centering{\begin{minipage}{12.2truecm}\footnotesize\baselineskip=12pt\ 
noindent
	\centerline{\footnotesize ABSTRACT}\vspace*{0.3cm}
	\parindent=0pt #1
	\end{minipage}}\par}}

\renewenvironment{thebibliography}[1]
	{\begin{list}{\arabic{enumi}.}
	{\usecounter{enumi}\setlength{\parsep}{0pt}
\setlength{\leftmargin 1.25cm}{\rightmargin 0pt}
	 \setlength{\itemsep}{0pt} \settowidth
	{\labelwidth}{#1.}\sloppy}}{\end{list}}

\newcounter{itemlistc}
\newcounter{romanlistc}
\newcounter{alphlistc}
\newcounter{arabiclistc}

\newcommand{\fcaption}[1]{
        \refstepcounter{figure}
        \setbox\@tempboxa = \hbox{\footnotesize Fig.~\thefigure. #1}
        \ifdim \wd\@tempboxa > 6in
           {\begin{center}
        \parbox{6in}{\footnotesize\baselineskip=12pt Fig.~\thefigure. #1}
            \end{center}}
        \else
             {\begin{center}
             {\footnotesize Fig.~\thefigure. #1}
              \end{center}}
        \fi}

\newcommand{\tcaption}[1]{
        \refstepcounter{table}
        \setbox\@tempboxa = \hbox{\footnotesize Table~\thetable. #1}
        \ifdim \wd\@tempboxa > 6in
           {\begin{center}
        \parbox{6in}{\footnotesize\baselineskip=12pt Table~\thetable. #1}
            \end{center}}
        \else
             {\begin{center}
             {\footnotesize Table~\thetable. #1}
              \end{center}}
        \fi}

\def\@citex[#1]#2{\if@filesw\immediate\write\@auxout
	{\string\citation{#2}}\fi
\def\@citea{}\@cite{\@for\@citeb:=#2\do
	{\@citea\def\@citea{,}\@ifundefined
	{b@\@citeb}{{\bf ?}\@warning
	{Citation `\@citeb' on page \thepage \space undefined}}
	{\csname b@\@citeb\endcsname}}}{#1}}

\newif\if@cghi
\def\cite{\@cghitrue\@ifnextchar [{\@tempswatrue
	\@citex}{\@tempswafalse\@citex[]}}
\def\citelow{\@cghifalse\@ifnextchar [{\@tempswatrue
	\@citex}{\@tempswafalse\@citex[]}}
\def\@cite#1#2{{$\null^{#1}$\if@tempswa\typeout
	{IJCGA warning: optional citation argument
	ignored: `#2'} \fi}}

 1
 1
 1

\font\ninerm=cmr9

\def\l{\langle}
\def\r{\rangle}

\begin{document}

%\runninghead
%{Universal Finite-Size-Scaling Functions}
%{Universal Finite-Size-Scaling Functions}

\normalsize %\textlineskip
\thispagestyle{empty}
\setcounter{page}{1}

%\copyrightheading{} %{Vol. 0, No. 0 (1995) 000--000}
\vspace*{0.88truein}

%\fpage{1}
\centerline{\bf UNIVERSAL FINITE-SIZE-SCALING FUNCTIONS}

\vspace*{0.37truein}
\centerline{\footnotesize YUTAKA OKABE\footnote{E-mail:
okabe@phys.metro-u.ac.jp}}
\vspace*{0.015truein}
\centerline{\footnotesize\it Department of Physics, Tokyo Metropolitan
University, Tokyo 192-03, Japan}
\vspace*{10pt}
\centerline{\normalsize and}
\vspace*{10pt}
\centerline{\footnotesize MACOTO KIKUCHI\footnote
{E-mail: kikuchi@phys.sci.osaka-u.ac.jp}}
\vspace*{0.015truein}
\centerline{\footnotesize\it Department of Physics, Osaka University,
Toyonaka 560, Japan}
\vspace*{0.225truein}
%\publisher{(received date)}{(revised date)}
%\pub{(received date)}

\vspace*{0.21truein}
\abstracts{
The idea of universal finite-size-scaling functions of the Ising model
is tested by Monte Carlo simulations for various lattices.  Not only
regular lattices such as the square lattice but quasiperiodic lattices
such as the Penrose lattice are treated.
We show that the finite-size-scaling functions of the order parameter
for various lattices are collapsed on a single curve by choosing
two nonuniversal scaling metric factors.  We extend the idea of the universal
finite-size-scaling functions to the order-parameter distribution
function.  We pay attention to the effects of boundary conditions.
}{}{}

\vspace*{10pt}
%\keywords{Universal Finite-Size-Scaling Function; Ising Model;
%Order-Parameter Probability Distribution Function.}

\vspace*{1pt}%\textlineskip

\section{Introduction}
\noindent
Quite recently, Hu {\it et al.}\cite{Hu95} have discussed the universal
scaling functions of the finite-size scaling, studying the two-dimensional
(2D) percolation problems.  They have shown that the data of the existence
probability and the percolation probability of bond and site percolation
on various lattices fall on the same universal scaling functions.
The idea of the universal finite-size-scaling functions was first
proposed by Privman and Fisher.\cite{Privman84}

In this article, the finite-size-scaling functions of the Ising model
are studied by Monte Carlo simulations for various lattices.  We deal
with not only regular lattices such as the square lattice, the triangular
lattice, {\it etc.}, but quasiperiodic lattices such as the Penrose lattice
and its dual lattice.  We show that the finite-size-scaling functions
of order parameter, $m$,  as a function of temperature, $T$,  for various
lattices are collapsed on the universal curve by choosing two scaling metric
factors.  Attention is paid to boundary conditions in connection with surface
effects.  The idea of the universal finite-size-scaling functions
is extended to the direction of external field, $H$.  In the phase plane
of temperature and field, we discuss a universal curved surface
for the finite-size scaling.  Generalizing the idea of the universal
finite-size-scaling functions, we argue the scaling of the order-parameter
distribution function.

\section{Universal scaling functions for the equation of state}
\noindent
We study the Ising model on various lattices by Monte Carlo simulations.
The simulation is performed at the critical point, $T=T_c$ and $H=0$, and
all the information near the critical point is obtained with the trick
of a histogram method.\cite{Ferrenberg88}

The temperature dependence of the second moments of order parameter
for the square lattice and the Penrose lattice, as examples, are shown
in Fig.~1.  The data for various sizes with periodic and free boundary
conditions are plotted with a unit of $J$.
\begin{figure}[htbp]
\vspace*{13pt}
\centerline{\vbox{\hrule width 5cm height0.001pt}}
\vspace*{5.9cm}
\centerline{\vbox{\hrule width 5cm height0.001pt}}
\vspace*{13pt}
\fcaption{Temperature dependence of the order parameter for the square (sq)
lattice and the Penrose (Pen) lattice.  Both the data for periodic and free
boundary conditions are plotted.}
\end{figure}

One comment should be made here on the Penrose lattice.
We use the `periodic' Penrose lattice, which has been previously
studied by Okabe and Niizeki.\cite{Okabe88}  The irrational golden
ratio, $(1+\sqrt{5})/2$, is approximated by a consecutive pair of
the Fibonacci numbers.  Thus, the number of spins, $N$, of
the `periodic' Penrose lattice are restricted to 76, 199, 521,
1364, and so on.  The unit cell of the present system is represented
by a large rhombus.

\begin{figure}[htbp]
\vspace*{13pt}
\centerline{\vbox{\hrule width 5cm height0.001pt}}
\vspace*{5.9cm}
\centerline{\vbox{\hrule width 5cm height0.001pt}}
\vspace*{13pt}
\fcaption{Universal finite-size-scaling plots of the order parameter.
The values of the metric factors, $C_1$ and $C_2$ are given in Table 1.}
\end{figure}
The finite-size-scaling plots for the order parameter are given in Fig.~2.
Following the standard finite-size scaling, the temperature and the order
parameter are scaled as $t L^{1/\nu}$ and $\l m^2 \r L^{2\beta/\nu}$,
where $t = T-T_c$ with $T_c$ being the critical temperature,
and $1/\nu$ and $\beta/\nu$ are the thermal and magnetic critical
exponents, respectively.  For the 2D Ising model, it is well known
that $\nu=1$ and $\beta=1/8$.  The linear dimension of the system
is denoted by $L$, and for the Penrose lattice,
$L$ is determined by $L=\sqrt{N}$.  In Fig.~2, we have
introduced two metric factors, $C_1$ and $C_2$.
Thus, the scaling form of the equation reads
\begin{equation}
 C_2^2 \l m^2 \r L^{2\beta/\nu}
     = f(C_1 t L^{1/\nu}) ,
\end{equation}
We have chosen $C_1 = C_2 = 1$ for the square lattice as a standard.
The metric factors of the Penrose lattice are determined
so as to get the best fit.  The data for other lattices are
also collapsed on the same curves with choosing
appropriate metric factors.  The estimated metric factors
together with $T_c$ are tabulated in Table~1.  The data for
the triangular lattice, the honeycomb lattice, and the dual Penrose
lattices are also shown in Table~1.
It should be emphasized that the obtained $C_1$ and $C_2$ are the same
for the periodic and free boundary conditions.
\begin{table}[htbp]
\tcaption{Metric factors for the Ising model on various lattices.}
\vspace{\baselineskip}
\begin{center}
\begin{tabular}{*{2}{|@{\quad} l}*{3}{@{\quad} l} |}
\hline
lattice & $T_c$ & $C_1$ & $C_2$ & $C_3$\\
\hline
square & 2.269$\dots$ & 1 & 1 & 1 \\
Penrose & 2.393$\pm$0.002 & 0.86$\pm$0.02 & 1.03$\pm$0.02
        & 0.95$\pm$0.02\\
dual Penrose  & 2.150$\pm$0.002 & 1.04$\pm$0.02 & 0.97$\pm$0.02
              & 1.05$\pm$0.02\\
triangular & 3.641$\dots$ & 0.60$\pm$0.02 & 1.02$\pm$0.02 & 0.98$\pm$0.02\\
honeycomb & 1.519$\dots$ & 1.50$\pm$0.03 & 0.98$\pm$0.02 & 1.02$\pm$0.02\\
\hline
\end{tabular}
\end{center}
\end{table}

The Binder parameter,\cite{Binder81}
\begin{equation}
g =  \frac{1}{2} \ ( 3 - \frac{\l m^4 \r}{\l m^2 \r^2} ) ,
\end{equation}
is often used in the finite-size-scaling analysis of critical phenomena.
The raw data for the Binder parameter are shown in Fig.~3 for the square
lattice and the Penrose lattice.
\begin{figure}[htbp]
\vspace*{13pt}
\centerline{\vbox{\hrule width 5cm height0.001pt}}
\vspace*{5.9cm}
\centerline{\vbox{\hrule width 5cm height0.001pt}}
\vspace*{13pt}
\fcaption{Temperature dependence of the Binder parameter for the square (sq)
lattice and the Penrose (Pen) lattice.  Both the data for periodic and free
boundary conditions are plotted.}
\end{figure}

The universal finite-size-scaling plots for the Binder parameter are
given in Fig.~4.  Here the same metric factor $C_1$ is used as that
in Fig.~2.
\begin{figure}[htbp]
\vspace*{13pt}
\centerline{\vbox{\hrule width 5cm height0.001pt}}
\vspace*{5.9cm}
\centerline{\vbox{\hrule width 5cm height0.001pt}}
\vspace*{13pt}
\fcaption{Universal finite-size-scaling plots of the Binder parameter.}
\end{figure}
Thus, we have shown that the finite-size-scaling plot of the Binder
parameter becomes universal among various lattices by choosing
appropriate metric factors.  However, the Binder parameter strongly
depends upon boundary conditions.  It means that the Binder parameter
reflects the finite-size effects due to fluctuations.
The universality of the Binder parameter at the critical point has been
pointed out by Bruce.\cite{Bruce85}  Moreover, Kamieniarz and
Bl\"ote\cite{Kamieniarz93} have argued the aspect-ratio dependence of the
universal value of the Binder parameter for rectangular systems
using a transfer-matrix technique.

Next turn to the system with an external field.  The second-order
cumulants of the order parameter, $\l m^2 \r - \l m \r^2$,
for the square lattice and the Penrose lattice are shown in Fig.~5.
The system size is fixed for each lattice but the data
for various $h = H/T$ are plotted.  Here, we restrict ourselves
to the case of the periodic boundary conditions
to escape from the complication of the figure.
\begin{figure}[htbp]
\vspace*{13pt}
\centerline{\vbox{\hrule width 5cm height0.001pt}}
\vspace*{5.9cm}
\centerline{\vbox{\hrule width 5cm height0.001pt}}
\vspace*{13pt}
\fcaption{Temperature dependence of the second-order cumulant of
the order parameter for various external fields, $h=H/T$
with the periodic boundary conditions.}
\end{figure}

Let us try the universal finite-size-scaling plot,
\begin{equation}
 C_2^2 (\l m^2 \r - \l m \r^2) L^{2\beta/\nu}
     = f_2 (C_1 t L^{1/\nu}, C_3 h L^{\beta \delta/\nu}) .
\end{equation}
Here, $\beta \delta$ is the so-called gap exponent, and $\delta = 15$ for
the 2D Ising model.  We show the finite-size-scaling plot in Fig.~6,
where $C_1$ and $C_2$ are the same as before.  We have also introduced
a new numeric factor $C_3$.  In Fig.~6, the data for various sizes at
different temperatures with different fields fall on a universal
curved surface.
\begin{figure}[htbp]
\vspace*{13pt}
\centerline{\vbox{\hrule width 5cm height0.001pt}}
\vspace*{5.9cm}
\centerline{\vbox{\hrule width 5cm height0.001pt}}
\vspace*{13pt}
\fcaption{Universal finite-size-scaling plots of the second-order cumulant
of the order parameter.  For the square lattice, the data of
the system sizes $L$=24, 40 and 64 are shown.  For the Penrose lattice,
those of the sizes $N$=521, 1364 and 3571 are shown.
The numeric factors $C_1$ and $C_2$ are the same as in Figs.~2 and 4.
The values of the new numeric factor $C_3$ are given in Table~1.}
\end{figure}

In this way, we have shown the universal finite-size-scaling plot for the
order parameter in the phase plane of temperature and field.
It is a generalization of the scaling equation of state for the bulk
system.\cite{Widom75}

\section{Finite-size scaling of order-parameter distribution function}
\noindent
In this section, we extend the idea of the universal finite-size
scaling to the probability distribution function of the order parameter,
$P(m)$.  The finite-size scaling of the distribution function has been
frequently used in the analysis of Monte Carlo data, for example, for
the spin-glass problem.\cite{Bhatt85}

\begin{figure}[htbp]
\vspace*{13pt}
\centerline{\vbox{\hrule width 5cm height0.001pt}}
\vspace*{5.9cm}
\centerline{\vbox{\hrule width 5cm height0.001pt}}
\vspace*{13pt}
\fcaption{Order-parameter probability distribution function $P(m)$
for the square lattice and the Penrose lattice, where temperature and
field are chosen under the conditions, $C_1 t L = 1.0$ and
$C_3 h L^{1.875} = 0.8$.}
\end{figure}
In Fig.~7, we show one example of the data for $P(m)$ off the critical point.
Here, the temperature and the field are chosen under the conditions,
$C_1 t L = 1.0$ and $C_3 h L^{1.875} = 0.8$, for the systems of
different lattices with various sizes.  The numeric factors $C_1$ and $C_3$
are those determined before.  Figure~8 shows the universal
finite-size-scaling plot of the order-parameter distribution function;
all the data fall on the same universal curve.
\begin{equation}
P(m; T, h; L) =  C_2 L^{\beta/\nu} P_2(C_2 m L^{\beta/\nu};
                 C_1 t L^{1/\nu}, C_3 h L^{\beta \delta/\nu}) .
\end{equation}
The equation (5) is the general universal finite-size-scaling function
of the distribution function.
\begin{figure}[htbp]
\vspace*{13pt}
\centerline{\vbox{\hrule width 5cm height0.001pt}}
\vspace*{5.9cm}
\centerline{\vbox{\hrule width 5cm height0.001pt}}
\vspace*{13pt}
\fcaption{Universal finite-size-scaling plot of the distribution function
of the order parameter.}
\end{figure}

\section{Summary and Discussions}
\noindent
We have shown that the equations of state of 2D Ising model
for various lattices are described by the universal
finite-size-scaling function with a small number of
numeric factors.  The above statement holds not only for the scaling
along the temperature axis but in the whole $t-h$ plane in the
finite-size-scaling critical region.

We have introduced three numeric factors, $C_1$, $C_2$ and $C_3$.
However, all three numeric factors are not independent.  The relation
$C_3 = C_2^{-1}$ is expected from the scaling argument.  Actually,
this relation appears to hold from Table 1 within statistical errors.
Moreover, these numeric factors are related to the relative ratio of
the critical amplitudes\cite{Privman91} of bulk systems for
different lattices.  The detailed analysis of this problem will
be discussed elsewhere.\cite{Okabe95}

The universal finite-size-scaling functions depend on boundary conditions.
Actually, the difference of the order parameter between periodic and free
boundary conditions is related to the so-called surface magnetization,
$m_s$, in the study of surface effects in critical
phenomena;\cite{Binder83}
\begin{equation}
  m_s = [m_{{\rm free}} - m_{{\rm periodic}}]/L .
\end{equation}
We should also note that the finite-size-scaling functions are
also sensitive to the lattice anisotropy, which will be left to
a future study.

We have generalized the concept of the universal finite-size-scaling
function for the order-parameter probability distribution;
we have shown that the probability distribution function is described
by the general universal finite-size-scaling function.

Finally, we should make a comment.  Through the present study,
as a by-product, we have shown again that the Ising model in quasicrystals
belongs to the same universality class as that of regular lattices.

\nonumsection{Acknowledgments}
\noindent
This work is supported by a Grant-in-Aid for General Scientific Research
from the Ministry of Education, Science and Culture, Japan.

\nonumsection{References}

\end{document}